\documentclass[conference,a4paper]{IEEEtran}
\IEEEoverridecommandlockouts
\usepackage{cite}
\usepackage{amsmath,amssymb,amsfonts}
\usepackage{algorithmic}
\usepackage{graphicx}
\usepackage{textcomp}
\usepackage{soul}
\usepackage{xcolor}
\sethlcolor{yellow}

\def\BibTeX{{\rm B\kern-.05em{\sc i\kern-.025em b}\kern-.08em
    T\kern-.1667em\lower.7ex\hbox{E}\kern-.125emX}}

\begin{document}

\title{Deepfake Image Generation for Improved Brain Tumor Segmentation\\}

\author{\IEEEauthorblockN{Roa'a Al-Emaryeen$^{1}$, Sara Al-Nahhas$^{1}$, Fatima Himour$^{2}$, Waleed Mahafza$^{3}$, and\hspace{0.1cm}Omar Al-Kadi$^{1*}$}
\IEEEauthorblockA{$^{1}$King Abdullah II School for Information Technology, University of Jordan, Amman 11942, Jordan \\$^{2}$Faculty of Information Technology, Zarqa University, Zarqa 13110, Jordan \\$^{3}$Department of Diagnostic Radiology, Jordan University Hospital, Amman 11942, Jordan \\ $^{*}$\tt\small {o.alkadi@ju.edu.jo}}\\
}

\maketitle

\begin{abstract}

As the world progresses in technology and health, awareness of disease by revealing asymptomatic signs improves. 
It is important to detect and treat tumors in early stage as it can be life-threatening. Computer-aided technologies are used to overcome lingering limitations facing disease diagnosis, while brain tumor segmentation remains a difficult process, especially when multi-modality data is involved. This is mainly attributed to ineffective training due to lack of data and corresponding labelling. This work investigates the feasibility of employing deepfake image generation for effective brain tumor segmentation. To this end, a Generative Adversarial Network was used for image-to-image translation for increasing dataset size, followed by image segmentation using a U-Net-based convolutional neural network trained with deepfake images. Performance of the proposed approach is compared with ground truth of four publicly available datasets. Results show improved performance in terms of image segmentation quality metrics, and could potentially assist when training with limited data.

\end{abstract}


\section{Introduction}
Brain cancer is considered a life-threatening disease that affects individuals of all ages. According to the World Health Organization (WHO), in 2020, 308,102 new cases of brain and central nervous system cancer occurred (in both sexes and all ages) and 251,329 deaths occurred (also in both sexes and all ages) \cite{b1}. In order to diagnose brain cancer, imaging techniques such as magnetic resonance imaging (MRI) and computed tomography (CT) scans are commonly used. These imaging modalities provide detailed brain structure images for identifying tumors and monitoring their development.

Over the past decade, there were tremendous efforts devoted to computer-aided technologies to solve classification and segmentation tasks in the medical field, particularly improving efficacy and reliability of accuracy and speed. However, the growing interest in brain tumor segmentation field is accompanied by functional limitations. One of the major limitations facing researchers is the lack of publicly available datasets that assist with classification and segmentation tasks in the medical field. Additionally, deep learning models require long training time from scratch, which can be a significant drawback, especially when training multiple models or fine-tuning model hyperparameters for a new task. This apart from training deep learning models is considered computationally intensive and requires powerful GPU-accelerated computing capabilities. As a result, training machine learning models from scratch can have negative environmental impacts. Massive computational power and storage needed for training large deep learning models entails energy consumption, and hence contributes to the increase of the carbon footprint.

The reuse of a pre-trained model on a new problem, known as \textit{transfer learning} \cite{b7}, or by creating modified copies of data as in \textit{data augmentation} \cite{b2intro}‌, and reducing execution time as in multi-thread \textit{parallel processing} \cite{b3intro} can assist with partially tackling the aforementioned challenges. Although these techniques reduce the amount of data or time required for training, they still require to be fed with sufficient data as an example of the new problem domain, which is unfortunately scarce in the medical imaging domain. Besides, collecting data directly from hospitals and clinics is considered tedious, time-consuming and requires ethical approval. Another approach would be to artificially generate sufficient deepfake images -- by transfer learning -- to properly train the data hungry deep learning models.

\begin{figure}[htbp]
\centerline{\includegraphics[height=5cm, width=9cm]{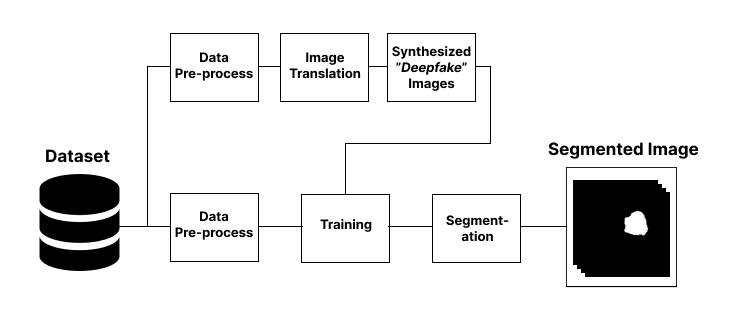}}
\caption{The primary procedures involved in this work. The initial step involves pre-processing the dataset, which is then followed by generating deepfake images through image translation. These generated images, in combination with a subset of the original dataset, are utilized to train the U-Net model. The resulting output of the U-Net is a segmented image, which is subsequently used for further analysis.}
\label{figure_pipeline}
\end{figure}

This work, described in Fig.\ref{figure_pipeline}, aims to investigate the feasibility of deepfake images in  deep learning model training for effective brain tumor segmentation. Findings of this research will contribute to a better understanding of the potential usefulness of deepfake images in improving diagnosis, and eventually the treatment of brain cancer. 

\section{Related Work}
Image segmentation is known as image partitioning into multiple parts or regions, often based on the pixel characteristics of an image. In this case, brain tumor segmentation is the process of separating abnormal tissues from normal brain tissues.  

Chang \cite{b2} proposed a Fully Convolutional Residual Neural Network (FCR-NN) based on linear identity mapping, a simple medical image segmentation method. The FCR-NN combines optimization gains from residual identity mappings with a fully convolutional architecture for image segmentation that efficiently accounts for both low- and high-level image features. The proposed model uses two different networks: the first to segment the entire tumor, and the other to segment sub-region tissues. The proposed model demonstrated improved performance with complete tumor, core tumor, and enhancing tumor validation Dice scores of 0.87, 0.81 and 0.72 respectively.

Zeineldin et al. \cite{b3} proposed a modular decoupling framework consisting of two main parts based on an encoding and decoding relationship, with spatial information extraction CNN for the encoder part. The semantic map resulting from the encoder is inserted into the decoder section to obtain a full-resolution probability map. A modified U-Net architecture was used with different CNN models, such as ResNet, DenseNet, and NASNet. The Dice and Hausdorff distance scores of the obtained segmentation results were 0.81 to 0.84 and 9.80 to 19.70, respectively.

For segmentation tasks, Tripathi et al. \cite{b4} proposed an OKM method. The OKM method is primarily a synthesis of two fundamental principles, Otsu thresholding and $k$-means clustering. The job includes segmenting the tumor and its components, namely necrosis, edema, and tumor enhancement. The results of the proposed approach were compared with the ground truth included in the BraTS dataset. The Dice coefficient score was 0.91 for 70 image slices. Also, Munir et al. \cite{b5} proposed a 2D-U-Net model based on convolution neural networks that was trained, validated and tested on the BraTS2019 brain tumor MRI dataset. Data augmentation was used to improve performance. An average Dice coefficient of 0.97 was achieved. 

In image generation, Shin et al. \cite{b29} proposes a context-aware generative adversarial network (CA-GAN) for synthesizing medical images, such as MRI and CT scans. The authors incorporate contextual information into the GAN architecture, allowing the generator to generate images that are consistent with the anatomical structure of the target region. The authors show that their CA-GAN approach outperforms other state-of-the-art methods on several medical image datasets and can generate high-quality images with clinically relevant features. This work is particularly important as it demonstrates the potential for GAN-based medical image generation to support clinical applications such as training deep learning models and developing new diagnostic tools.

Zhou et al. \cite{b30} propose a GAN-based approach for unsupervised liver lesion segmentation in CT scans. The authors train a GAN to generate synthetic CT images with and without liver lesions and use the discriminator network to distinguish between real and synthetic images. The authors then use the generator network to synthesize additional CT images with liver lesions and use these synthetic images to train a segmentation model. The authors show that their approach achieves competitive performance on several benchmark liver lesion segmentation datasets and can generate high-quality synthetic images that accurately capture the complex and variable nature of liver lesions. This work is important as it demonstrates the potential for GAN-based medical image generation to support unsupervised learning tasks, which could be particularly useful in settings where labeled medical images are scarce.

\section{Methodology}
In this section, dataset size will be artificially increased by generating deepfake images -- without the classical data augmentation approach, then the impact of the deepfake images by evaluating the performance and accuracy of the segmentation will be demonstrated. 

\subsection{Transfer Learning}
Transfer learning is a machine learning technique used to improve the performance of a model trained on a different task by using a pre-trained model as a starting point and fine-tuning using a smaller dataset. It is particularly useful in situations where the amount of labeled data is limited or the target task is related to the source task \cite{b6}, \cite{b7}. It is a powerful technique that can be applied to medical image translation tasks \cite{b8}.  In this work, it will be utilized in deepfake image generation and segmentation.

\subsection{Deepfake Image Generation} 
Medical image translation refers to the process of converting an image from one domain or modality to another, such as converting an MRI image to a CT image or vice versa. The goal of medical image translation is to generate images that are similar in appearance and information content to the original images. One of the main advantages of using transfer learning in medical image translation is that it can significantly reduce the amount of labeled data required for training \cite{b9}. 

\begin{figure}[htbp]
\centerline{\includegraphics[height=4cm, width=8cm]{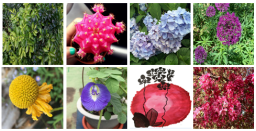}}
\caption{Sample images in iphone2dslr flower pre-trained model.}
\label{figure1}
\end{figure}

In this paper, faked images are generated using the general CycleGAN framework \cite{b10, b26} and the iphone2dslr\texttt{\_}flower pre-trained model through image-to-image translation \cite{b11} using a generative model such as a GAN \cite{b12}. The iphone2dslr\texttt{\_}flower pre-trained model is a generative model that has been trained on a large dataset of images. Fig.\ref{figure1} shows some of the flowers that are used in iphone2dslr model.

The model can learn image features and patterns in a particular domain for generating new ones in a different domain. The process includes loading the pre-trained model, preparing the input image, using the pre-trained model to generate fake images, post-processing, and evaluating the generated image using metrics, such as mean square error and structural similarity index measure \cite{b13}. It is worth mentioning that the iphone2dslr\texttt{\_}flower model is a conditional generative model that can generate realistic flower images from a given input image, it is not directly designed for medical image translation, but with some adjustments and fine-tuning, it could be used to generate images that are similar in appearance and information content to the original brain images.

The cycleGAN is a type of generative adversarial network (GAN) that is trained to map images from one domain to another. It is composed of two generators, $G$ and $F$, and two discriminators, $D_{x}$ and $D_{y}$. The generators are trained to transform images from domain $X$ to domain $Y$ and vice versa, while the discriminators are trained to distinguish between real images from their respective domains and fake images generated by the generators. The loss function for the cycleGAN includes two components: the adversarial loss and the cycle consistency loss.

\subsection{Medical Image Segmentation}
Deep neural networks have shown remarkable results for various challenging image segmentation and classification problems. Nonetheless, training deep neural networks is still difficult because of the limited training data. One of the proven methods to overcome this problem is to initialize the weights of a convolutional network that has been pre-trained on a large dataset, so to improve performance for a specific task such that a limited number of training data is available \cite{b14}.

\begin{figure}[htbp]
\centerline{\includegraphics[height=6cm, width=10cm]{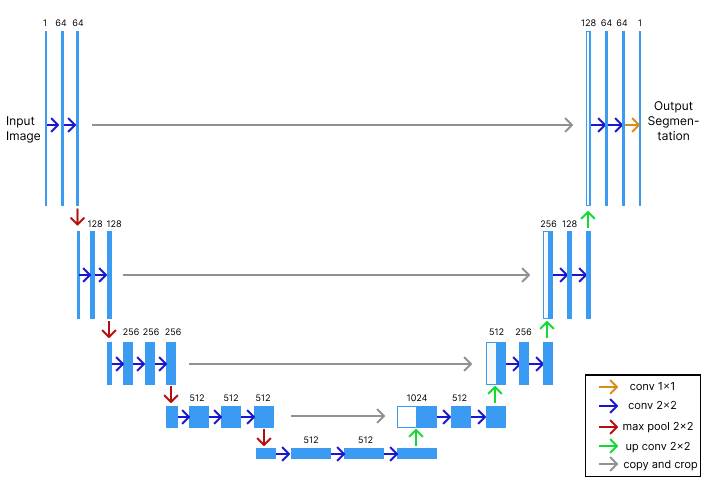}}
\caption{U-Net convolutional network architecture.}
\label{figure4}
\end{figure}

We evaluate in this work the segmentation performance of a U-Net architecture \cite{b15, b16} when trained on real and deepfake images. The architecture consists of a down-sampling (i.e. encoding) path, a bottleneck, and an up-sampling (i.e. decoding) path as shown in Fig. \ref{figure4}. In the down-sampling process, convolutional layers are used to increase the number of feature maps. Max pooling operation is also used to reduce the size of the feature maps. In the up-sampling process, deconvolutional layers are used to reduce the number of feature maps and increase the size of the feature maps. Convolutional layers are also used to reduce the number of feature maps that are concatenated from deconvolutional feature maps and feature maps from the encoding path.

\begin{figure}[htbp]
\centerline{\includegraphics[height=6cm, width=10cm]{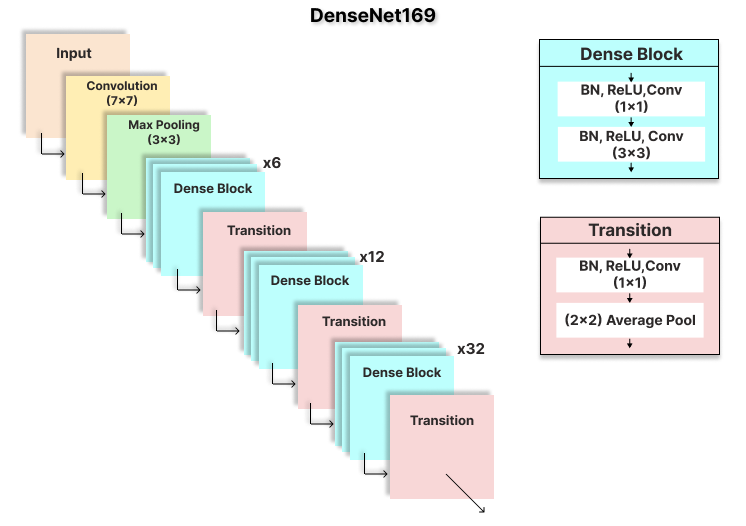}}
\caption{A pre-trained DenseNet-169 encoder with ImageNet weights.}
\label{figure6}
\end{figure}

The encoder used in this work, as shown in Fig. \ref{figure6}, is a densely connected convolutional network \cite{b17} that has been trained on the ImageNet dataset \cite{b18}. It is well known that DenseNet-169 model extracts rich and robust features from images, and pre-training on ImageNet dataset further improves its ability to extract relevant features for the segmentation task.

\section{Experimental results}

\begin{figure}[htbp]
\centerline{\includegraphics[height=4cm, width=8cm]{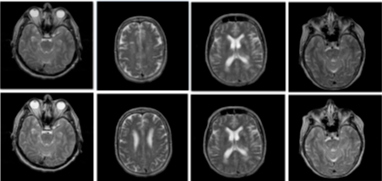}}
\caption{MR image generation: [Top row] real and [bottom row] generated 'deepfake' MR images.}
\label{figure2}
\end{figure}

Fig. \ref{figure2} present examples of MR images artificially generated through the image translation process. The top row of the figure displays the original images, while the bottom row shows the translated (i.e. deepfake) images. As can be seen, the generated images closely resemble the original images in terms of appearance and information content. 

\subsection{Experimental Datasets}
This work utilizes four publicly available datasets: Multimodal Brain Tumor Segmentation Challenge (BraTS) 2020 dataset \cite{b19, b20, b21, b22}; Unpaired MR-CT Brain Dataset for Unsupervised Image Translation \cite{b23}; Barin T1-weighted CE-MRI images \cite{b24}, and the IXI dataset \cite{b25} to train and evaluate architecture performance. 

The aspects of the used experimental datasets are as follows: \begin{itemize} \item The BraTS 2020 Challenge dataset \cite{b19, b20, b21, b22, b22_brats} contains 369 training and 125 validation multimodal brain magnetic resonance studies. Each study includes four MR sequences with same size of $240\times 240\times 155$: T1-weighted (T1), post-contrast T1-weighted (T1ce), T2-weighted (T2), and fluid-attenuated inversion recovery (Flair). For each study, experts annotated the enhancing tumor (ET), peritumoral edema (ED), and necrotic and non-enhancing tumor core (NCR/NET) on a voxel-by-voxel basis. In this work T1-weighted (T1) MR images were used.

\item The Unpaired MR-CT Brain Dataset \cite{b23} contains unpaired 2D MR and CT image slices from 20 patients. The dataset includes a total of 179 with size of $256\times 256$ 2D axial image slices for 20 patient volumes (90 MR and 89 CT 2D axial image slices). 

\item The Brain T1-weighted CE-MRI images dataset \cite{b24} contains 3064 T1-weighted contrast-enhanced images of 233 patients with three types of brain tumor: meningioma (708 slices), glioma (1426 slices), and pituitary tumor (930 slices). All MR images are 2D axial and $512\times512$ in size.

\item The IXI dataset \cite{b25} contains nearly 600 MR images of normal healthy subjects collected in three different hospitals in London. Various sequences such as T1, T2, proton density-weighted (PD), magnetic resonance angiography (MRA), and diffusion-weighted (DTI) images are provided.
\end{itemize}


During the pre-processing phase, the images are resized to $256\times256$ and normalized between 0 and 1. The number of images used from each dataset in this work are included in Table \ref{table:dataset_info}. The total number of images obtained and combined from the first four datasets is 5290 images and then are split into training and validation sets in an 80\%-20\% ratio.

\begin{table}[ht]
\caption{Total number of images for model training}\label{table:dataset_info}
\centering
\begin{tabular}{ccccc}
\hline
Dataset Name & Extracted Images\\
\hline
BraTS 2020  & 1529\\
Unpaired MR-CT & 270\\
Brain tumor & 491\\
IXI & 3000 &\\
Deepfake images & 174\\
\hline
\end{tabular}
\end{table}

\begin{figure}[htbp]
\centerline{\includegraphics[height=5cm, width=8cm]{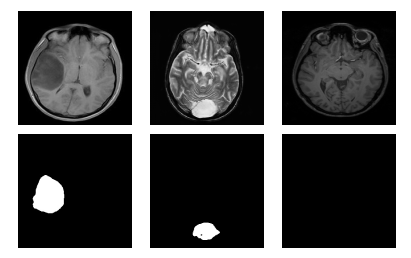}}
\caption{Segmentation based on enhancing training with MR deepfake images: [Top row] sample MR images and [bottom row] segmented tumor regions. Note last image is normal, therefore no segmentation output is shown.}
\label{figure5}
\end{figure}

The network was trained before (total of 5290 images) and after the inclusion of deepfake images (total of 5464 images). Training was set to 25 epochs, a batch size of 32, Dice loss as a loss function \cite{dice_loss}, and Adam optimizer \cite{b25_adam} with a learning rate of 0.0001 used in each training. The training was performed using local Jupyter Notebook, Python 3.9, and NVIDIA GeForce RTX 2060 SUPER GPU with 8 GB of memory. Fig. \ref{figure5} illustrates the input and output of using DenseNet169-UNet for brain tumor segmentation. 

In order to evaluate network performance, overlap-based and surface-based metrics were used. To evaluate the similarity/overlap between the ground truth and the predicted mask, the Dice similarity coefficient (DSC) and Jaccard similarity coefficient (JSC) metrics were utilized. The main difference between the DSC and JSC is how they penalize false positives and false negatives. False negatives are typically given greater weight by the DSC, but false positives are typically given more weight by the JSC.

The mean absolute distance (MAD) and Hausdorff distance (HD) metrics were used to evaluate the distance/error between two points in a metric space. The primary difference between MAD and HD is the distance measurement. MAD is the average distance between the two sets, whereas HD is the maximum distance between any point in one set and the nearest point in the other set. Therefore, both metrics were used to provide a more comprehensive evaluation of the segmentation performance.

The validation results of training the model with and without deepfake images are shown in Table \ref{table:validate}. 

\begin{table}[ht]
\caption{Validation results of training without ($T$) and with ($T_{DF}$) deepfake images}\label{table:validate}
\centering
\begin{tabular}{ccccc}
\hline\hline
& DSC & JSC & MAD & HD \\
\hline
T & 0.53 $\pm$ 0.28 & 0.41 $\pm$ 0.26 & 0.084 $\pm$ 0.088 & 0.27 $\pm$ 0.08\\
\hline
\textbf{$T_{DF}$} &  \textbf{0.59 $\pm$ 0.26} & \textbf{0.46 $\pm$ 0.25} & \textbf{0.061 $\pm$ 0.056} & \textbf{0.25$\pm$ 0.05}\\ 
\hline\hline
\end{tabular}
\end{table}

\section{Discussion and Conclusion}

\begin{figure}[htbp]
\centerline{\includegraphics[height=9cm, width=9cm]{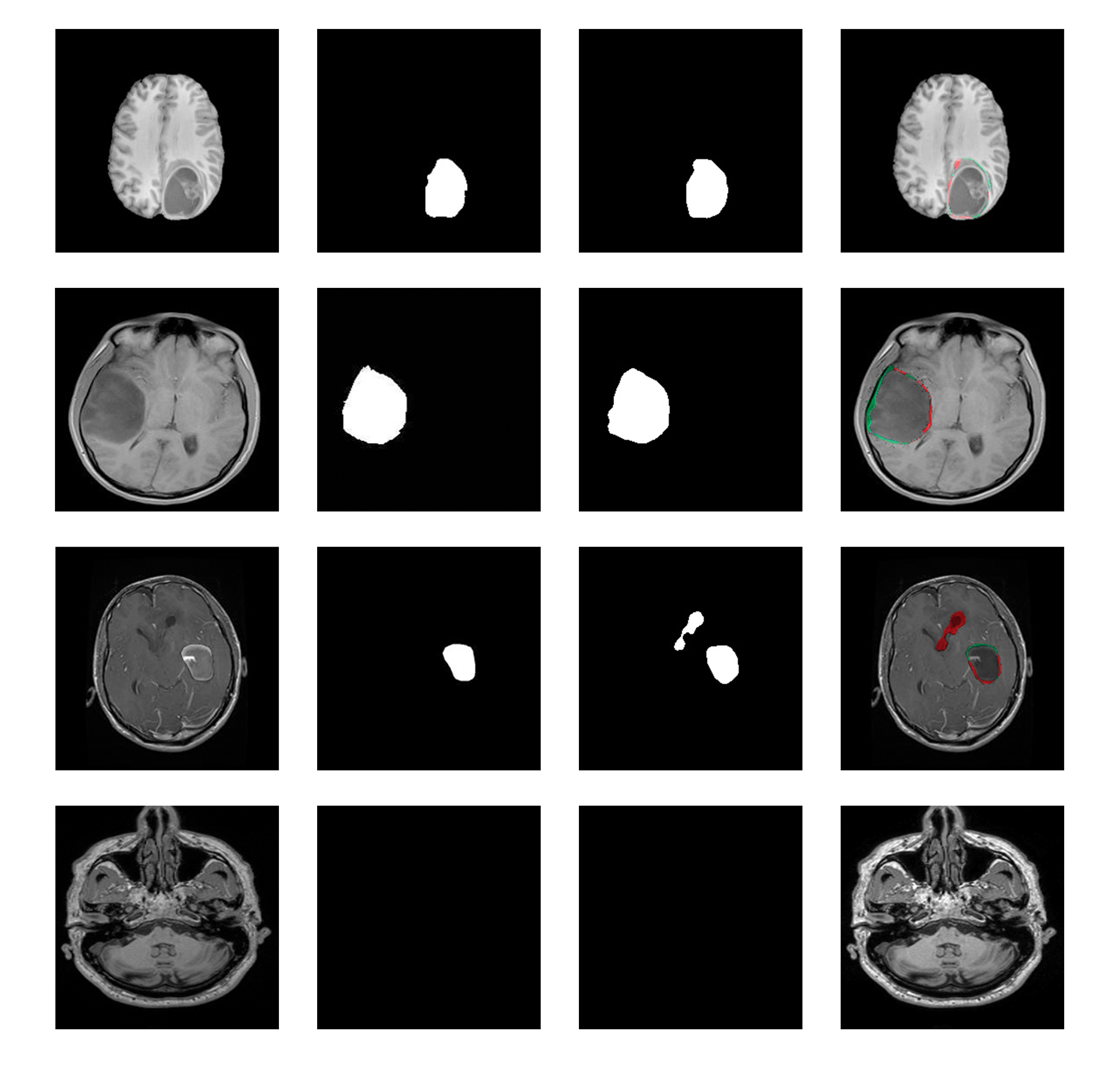}}
\caption{Results obtained without enhancing training with deepfake images. [left-right] Ground truth MR image, corresponding lesion mask, segmented lesion area, overlapping between lesion mask and segmented area on MR images, respectively (gray color represents true positive, green color for false negative, and red color shows false positive.}
\label{figure7}
\end{figure}

\begin{figure}[htbp]
\centerline{\includegraphics[height=12cm, width=9cm]{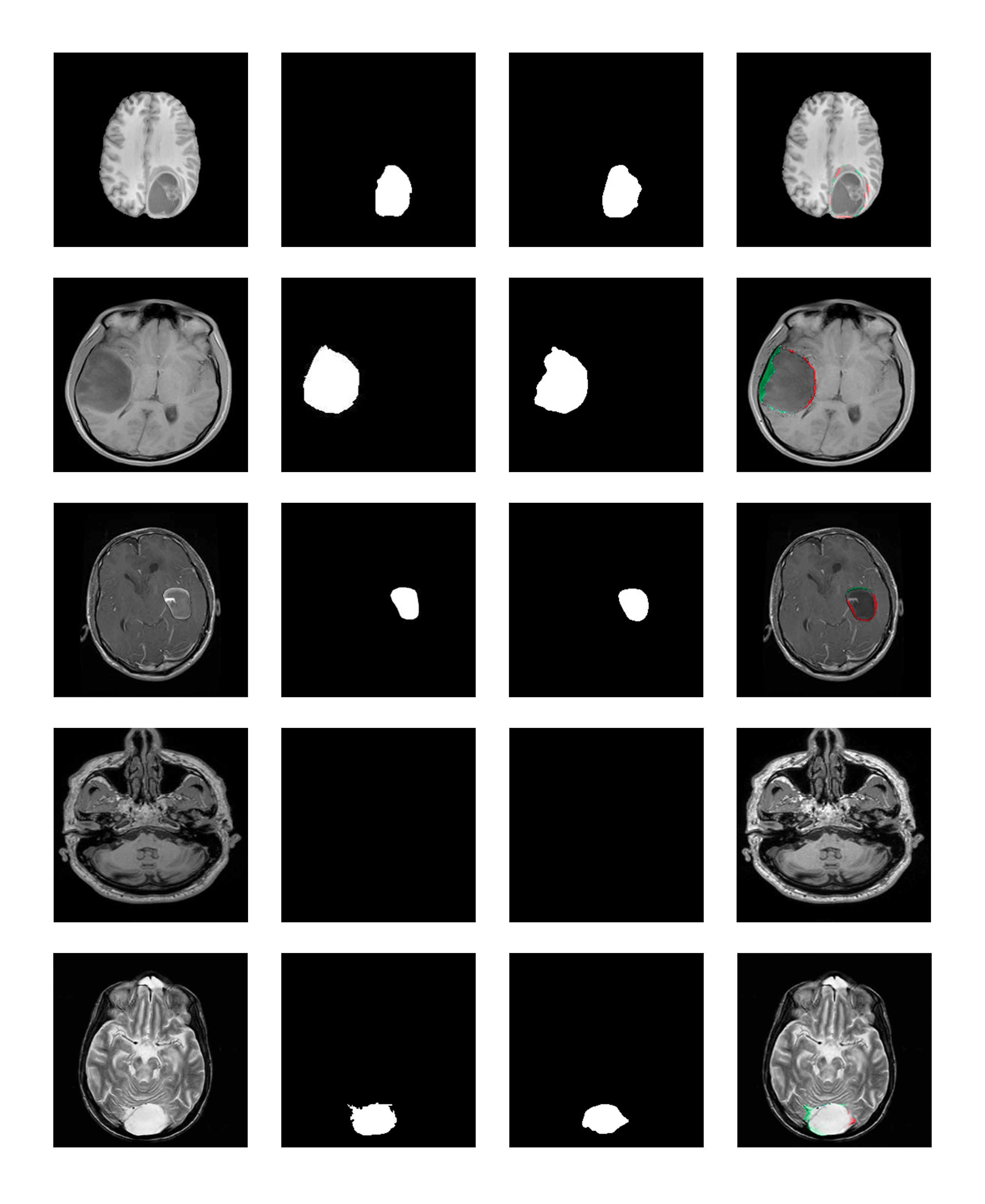}}
\caption{Enhancing training with deepfake images. [left-right] the ground truth MR image, corresponding lesion mask, segmented lesion area, overlapping between the lesion mask and segmented area on MR images, respectively (gray color represents true positive, green color for false negative, and red color shows false positive.}
\label{figure8}
\end{figure}

The network performance improved after using deefake images, as observed in Table \ref{table:validate}. Both Figs \ref{figure7} and \ref{figure8} visually illustrate the network performance before and after enhancing training with deepfake images.

The columns from left to right in Fig. \ref{figure7} represent the ground truth MR image, corresponding lesion mask, segmented lesion area without deepfake images, overlapping between the lesion mask and segmented area on MR images without deepfake training, respectively. In Fig. \ref{figure8} represents the ground truth MR image, corresponding lesion mask, segmented lesion area with deepfake images, overlapping between the lesion mask and segmented area on MR images with deepfake training, respectively.
 

To validate the performance of the network accurately, the same MR images are used in both Figs. \ref{figure7} and \ref{figure8}. There are fewer false positives regions in Fig. \ref{figure8}, which shows that after employing the deep-fake images, the network's performance improved and its ability to predict correct lesions increased.


Various medical tasks such as classification and segmentation require a vast number of images to be used in order to provide reliable results. In this work, the use of deepfake images to segment brain tumors improved the performance of the model compared to real images. These results suggest that deepfake images provide robust and stable performance and improve model generalization. 

This research has highlighted training with deepfake images as a promising approach, and holds potential as a valuable technique in enhancing medical image segmentation. However, it should be noted that there could be limitations, as deepfake data needs prior registration between image pairs, and performance was not investigated in case of small lesions or noise presence \cite{b27,b28}. Future work may include expanding the dataset size and investigate deepkafe image robustness under noisy conditions. Furthermore, further research can be done to validate the efficacy of this approach through large-scale clinical trials. To conclude, while this work demonstrates the good side of deepfake images in improving image segmentation, further clinical validation is required to fully realize its potential in real practice.

\section*{Acknowledgment}
This work was funded by The Jordan Scientific Research Support Fund (Grant no. ICT/1/03/2021). We would like to thank Dr. Isra'a Almallahi and Dr. Ruba Braik from Jordan University Hospital for assisting with annotating the deepfake images.

\end{document}